\documentclass[prl,twocolumn,showpacs,amsmath,amssymb]{revtex4}

\usepackage{graphicx}
\usepackage{dcolumn}
\usepackage{bm}

\begin{document}%


\title{Time-resolved Photo-Phonon Spectroscopy\\of Exchange Coupled
${Cr^{3+}}$ - Pairs in Ruby}

\author{A. M. Shegeda}
 \email{shegeda@kfti.knc.ru}
\author{V. N. Lisin}%
 \email{vlisin@kfti.knc.ru}
\affiliation{%
Zavoisky Physical\d Technical Institute of the Russian Academy of
Sciences, 10/7 Sibirsky Trakt Street, Kazan, 420029, Russia
}%
\date{\today}
\begin{abstract}
A rather simple method is used to detect at once both optical
absorption spectra and excited-states nonradiative transitions  in
0,03~at~${\%}$ and 0,16~at~${\%}$~ruby at temperature 2~K. The
technique utilizes a time-resolved bolometer detection of phonons,
generated by the excited-state nonradiative relaxation following
optical excitation with a pulsed tunable dye laser. The observed
excitation spectra of phonons well coincide with already known
absorption spectra of both single chromium ions and pairs of the
nearest neighbors. For the first time the fast ($\leqslant
0,5~{\mu}s$)resonant energy transfer from single chromium ions to
the fourth- nearest neighbors is observed directly. The new
strongly perturbed $Cr$-single-ion sites are observed.

\end{abstract}

\maketitle
\bibinfo

The used method can be referred, as photo-phonon spectroscopy. The
nonequilibrium phonons (NP), which are created after laser pulse
excitation due to radiationless relaxation, can be detected as
bolometer temperature changing in real time following laser
excitation, if a free length of the phonons is comparable to
distance up to a bolometer. Therefore, the bolometer signal
amplitude is determined by both the absorption cross section and
the radiationless relaxation intensity so both absorption spectra
and radiationless relaxation can be studied by this method. For
the first time this technique was proposed in \cite{rob77}. It is
a natural development of the photo-acoustic \cite{ros80} method.
Absorption spectra of ruby in the $R_{1}$~-~$R_{2}$ region were
observed in \cite{rob77} utilizing sinusoidally chopped light
excitation.  We use a pulse laser excitation and wider rang of the
light  wave lengthes then in \cite{rob77}. It has allowed us to
identify the lines of chromium ion pair and to deduct about the
fast resonant energy transfer from single ions to the fourth
nearest neighbors and about new single ion sites.
\begin{figure}
\includegraphics{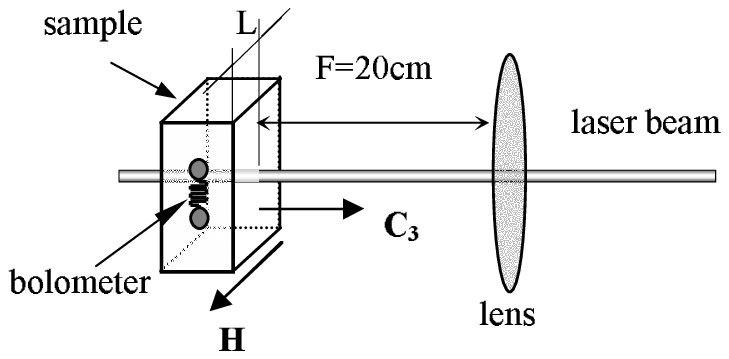}
\caption{\label{fig:plan} The experimental geometry. \\ \textbf{H}
and ${C_3}$ - directions of an exterior constant magnetic field
and an optical axis, L - the distance between a laser beam and a
bolometer (it is fixed in our experiments), F - focal distance.}
\end{figure}
The geometry of experiment is shown in Fig.~\ref{fig:plan}. The
dye (oxazin~1) tunable laser is used for optical excitation of
${Cr^{3+}}$ ions. It produces light pulses of $8~ns$ in duration
with a repetition frequency 12,5~Hz~and a spectral half-width ~
$0,1~\text{\AA}$ at any wave length in the range $6915~
\text{\AA}~\div~ 7100~ \text{\AA}$. The samples of a ruby have
${Cr^{3+}}$ concentration 0,03~at${\%}$ and 0,16~at${\%}$ and
thickness (along axis ${C_{3}}$) $10,0~mm$ and $5,0~mm$
accordingly. Linearly polarized laser radiation is transmitted
through a sample along the $C_3$-axis. The phonon pulse created at
passage of laser radiation through a sample are detected by a
superconductive Sn- or In-bolometer. The bolometer was evaporated
on one of lateral sides of a sample. It's shape is a meander and
area $3~mm^{2}$. Samples are immersed in liquid helium which is
pumped down to 2~K. The working point of a bolometer is tuned to
the inferior part of the linear site of superconductive transition
due to an external magnetic field. The bolometer signals are
amplified using a wide-band preamp with a small noise signal. Data
output is via stroboscopic oscilloscope. All processes of
measuring (detecting of a signal, its accumulation and processing,
change of a wave length of the laser) were automated.

It is necessary to note, that the laser wave length  can be
changed with a step, smaller 0,005~\AA, however the reflecting
monochromator, used by us, did not allow to establish the wave
length values with this accuracy. Precisely enough it was possible
to establish only the wave length relative change. Besides in the
literature there is no unambiguity in  wave length values of the
$R_{1}$ and $R_{2}$ lines of a ruby. Though some authors give its
to within the 0,01~\AA, the wave length value, for example,
$R_{1}$ lines at helium temperature in different works varies from
6933,6~\AA~to~ 6935,0~\AA~\cite{Kis69,engs73,pow72}. Therefore, in
order to prevent confusion at identification of lines, we further
considered, as well as in~\cite{Kis69} where the complete data on
absorption spectra and luminescence of the concentrated ruby are
cited most, $\nu(R_{1}) = 14418,52~ cm^{-1}$, i.e.,
$\lambda(R_{1})$ =6933,6~\AA.

\begin{figure}
\includegraphics{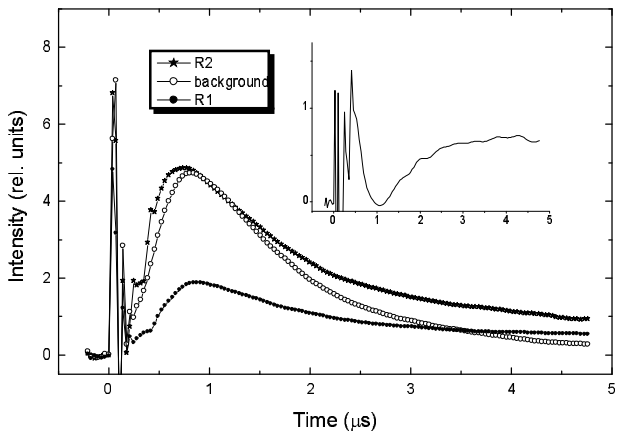}
\caption{\label{fig:time} Bolometer signals  in the ruby (
${0,16~at\%~ Cr^{3+}}$) versus the time after the laser pulse
excitation : Background -~nonresonance excitation ($\lambda
=6922~\text{\AA}$), R1 and R2 -~resonance excitation of the
$R_{1}$ and $R_{2}$ lines. The inset shows the difference between
the R2 and background phonon signals. Time - is delay time after
the laser pulse excitation.}
\end{figure}

\begin{figure}
\includegraphics{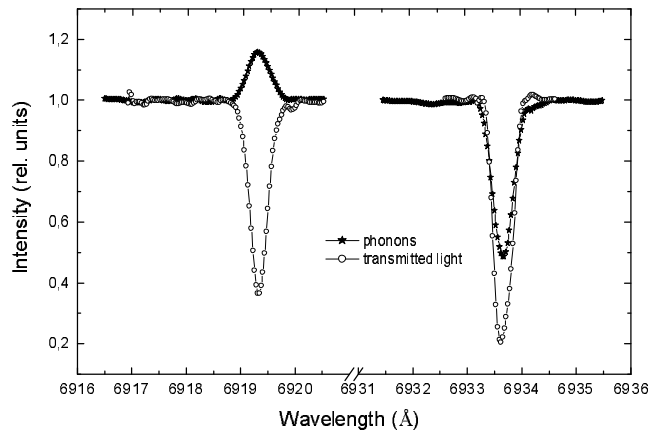}
\caption{\label{fig:low} Transmissivity  and phonon excitation
spectra near the $R_{1}$ and $R_{2}$ lines  in the low
concentrated ruby sample (${0,03~at\%~ Cr^{3+}}$). Transmissivity
curves are shifted downwards on 0,1 for presentation.}
\end{figure}

\begin{figure*}
\includegraphics{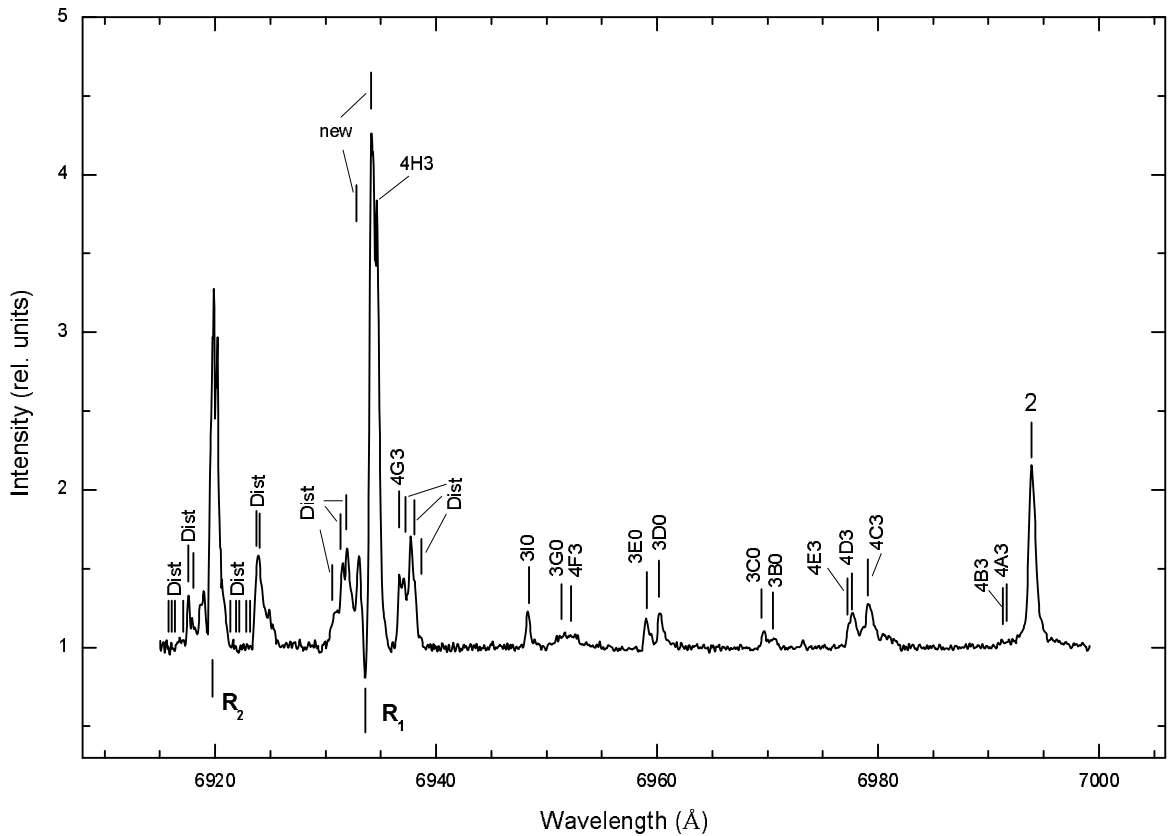}
\caption{\label{fig:specter} Phonon excitation spectra at
3~${\mu}s$ delay time after the laser pulse  in ${0,16~at\%~
Cr^{3+}}$~ruby at 2~K versus on the laser wave length.  The lines
show locations of the all absorption lines at the region
$6915~\AA~\div~ 7000~\AA$ ~found in ~\cite{Kis69} for temperature
4~K and polarization $\textbf{E}\bot C_3$. Labels as
in~\cite{Kis69}. The line $4H3$ is not mentioned in~\cite{Kis69},
the line $4F3$ is mentioned in ~\cite{Kis69} , but the transition
frequency value is not given so these values  are accepted equal
to the calculated values given in ~\cite{Kis69}: $\nu(4H3) =
14416,2~ cm^{-1}$ and $\nu(4F3) = 14379,9~ cm^{-1}$. The value of
frequency of transition $4G3$ is given as in~\cite{Kis69} for
polarization $\textbf{E}\parallel C_3$.}
\end{figure*}

\begin{figure}
\includegraphics{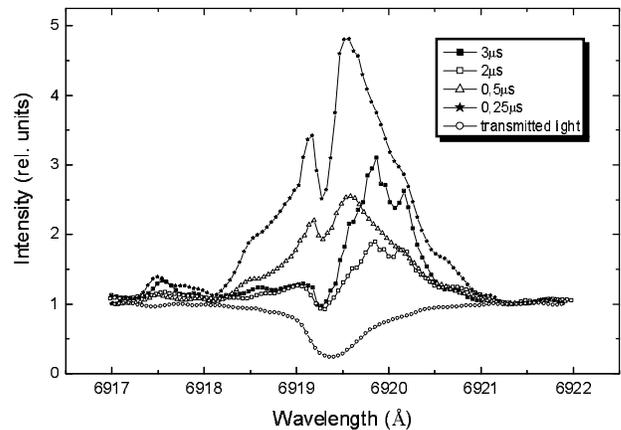}
\caption{\label{fig:R2} $R_{2}$-region of the laser pulse
transmissivity and the phonon excitation spectra of the
Fig.~\ref{fig:specter} at the different delay times after the
laser pulse input.}
\end{figure}

\begin{figure}
\includegraphics{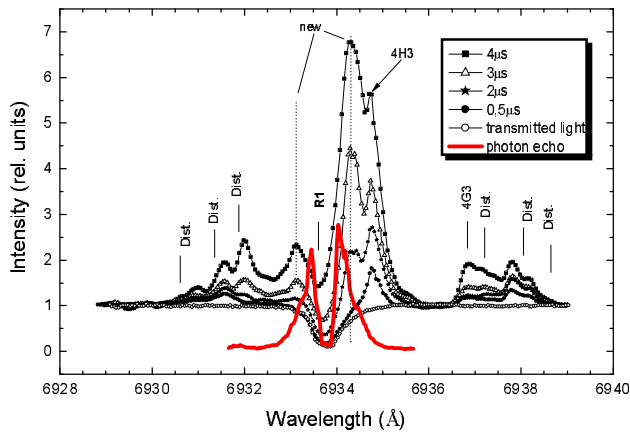}
\caption{\label{fig:R1} $R_{1}$-region of the phonon excitation
spectra of Fig.~\ref{fig:specter} at the different delay times
after the laser pulse input. Also the laser pulse transmissivity
and amplitude of the backward two pulse (delay time $30,0~ns$
between the pulses) photon echo near the $R_{1}$ line are shown.}
\end{figure}

When a sample is excited by a laser pulse, the bolometer detects a
phonon pulse after some delay time depending on the distance $L$
between the impact laser point and the bolometer ($L=2,5~mm$ for
results, shown in Figs.~\ref{fig:time}-\ref{fig:R1}).
Typical signals are shown in Fig.~\ref{fig:time}. The spike at the
time initial moment in this figure is because the bolometer
detects not only phonons, but also the light scattered in the
sample. The phonon pulses were observed at all the explored laser
wave length band, i.e., that is also and at nonresonance laser
excitation, which light frequency was differed from frequencies of
optical transitions both single ions as pairs of the nearest
neighbors. The reason of nonresonance laser generation of these
background phonons is not clear and demands separate studying.
During wave length scanning the amplitude of a bolometer signal is
measured at some fixed delay time after the laser pulse
excitation. The number of accumulation in  each point is equal 50.
Then the  specter   is normed on the background, because the dye
laser pulse intensity depends on wave length. Further, everywhere,
where we shall speak about a phonon excitation specter, the
specter received thus will be meant.

 At scanning of a wave length of the laser in a low concentrated ruby
  we observed change of the phonon signal only in the vicinity of the $R_1$
  and $R_2$ lines (see Fig.~\ref{fig:low}).
As you can see from this figure, the considerable decreasing of
the phonon signal is observed at resonant pumping of $R_1$ line.
Probably, it is caused by decreasing of background phonons
generation due to absorption of a laser pulse in volume of the
sample, giving to decreasing of the laser pulse intensity. Growth
of the phonon signal at the resonance with a $R_2$ line may be
explained by creation of additional $29~{cm}^{-1}$ phonons
generated by the excited-state nonradiative relaxation
$2\bar{A}\rightarrow \bar{E}$ as in~\cite{hu80} .

The most interesting results were received in the concentrated
ruby sample (0,16~at\% ${Cr^{3+}}$). In our work the transmission
spectrum was measured without use of a monochromator. The wave
length of the input laser is varied. The output laser pulse, past
through a sample, is detected by a photomultiplier. Rather low
peak stability of  dye laser pulses, plus usage of the high-speed
photomultiplier, importing additional noise, and also is essential
a smaller chromium ion concentration gave to that in a
transmission spectrum  only $R_1$ and $R_2$ lines were well
observed as you can see in Fig.~\ref{fig:low}. However, when we
begin to detect phonon signal on a bolometer, but not a
transmission specter, we have received unexpected interesting
results (see Fig.~\ref{fig:specter}): There is addition generation
of phonons at particular excitation wave lengthes.

In Fig.~\ref{fig:specter} all known absorption lines of chromium
single ions and pairs of the nearest neighbors are shown also. In
view of that spectral width of the laser in our experiment about
0.1 A, coincidence of the experimental values of lengths of waves,
at excitation on which is observed change of phonon signal, to
absorption lines of the nearest neighbors is quite satisfactory.
It is visible, that the lines, excitation of which is not
accompanying with nonradiative transitions ($4A3$, $4B3$), are not
observed by the method utilized in the given work. Therefore, it
is possible to count, that change of the phonon signal  is bound
to generation of phonons due to nonradiative transitions from
optical excited state on a metastable level.

Let's consider more in details the phonon excited spectra in area
$R_1$ and $R_2$ lines. As you can see in the inset of
Fig.~\ref{fig:time}, $R_2$-excited phonon signal differs under the
shape from a background phonon signal and has  the expressed
ballistic character. Indeed, at short times equal the times of
flight of ballistic longitudinal and transverse 29~cm$^{-1}$
phonons, the bolometer signal amplitude has peaks. The excitation
wave lengthes of these phonons peaks are near the transmissivity
dip ($\lambda$ =6919,26~\AA), as you can see in Fig.~\ref{fig:R2}.
At increasing of a delay time, these peaks decrease and  two new
excitation lines of phonons  with $\lambda =6919,85~\text{\AA}$
and $\lambda =6920,17~\text{\AA}$ become visible in the phonon
excitation spectra. These lines are in longwave area of $R_{2}$
region.

Even more interesting situation is observed in the $R_{1}$ region,
as you can see in Fig.~\ref{fig:R1}. Already at small delay times
about 0.5~${\mu}s$ there is well allowed line lower than a
$R_{1}$-level on 1,1 A in the phonon excitation specter. It is
known, that approximately here there should be the line of the
fourth nearest neighbors $4H3$ which is very difficult to resolve
by direct optical methods: the energy level experimental value is
not given in ~\cite{Kis69}, $\nu(4H3) = 14416,7~
cm^{-1}$~\cite{kan70,mont87}, $\nu(4H3) = \nu(R_1)-2
cm^{-1}$~\cite{selz77} that is $\nu(4H3) = 14416,52~ cm^{-1}$.
Just approximately in this place ($\nu = 14416,36~ cm^{-1}$,
$\lambda$ =6934,65~\AA.) the peak in the phonon excitation spectra
is observed at all delay times, as you can see in
Fig.~\ref{fig:specter} and Fig.~\ref{fig:R1}. Presence or absence
of the line located on frequency is little bit lower than
$R_1$-line, it is very important ~\cite{selz77, mont87, zh94} for
examination of mechanisms of migration of energy from single ions
to pairs. However intensities of phonon signals on transitions
$4H3$ and $4G3$ pairs are expected equal~\cite{mont87}, and in our
experiment strongly differ. Probably, it is so because at the
$4H3$-transition  frequency simultaneously with pairs the single
ions are excited, which fastly transfer energy to ones. It is
possible to tell, that for the first time the fast resonance
transfer of energy from single ions to pairs of the fourth-nearest
neighbors has been detected directly.

At last, you can see in Fig.~\ref{fig:specter} and
Fig.~\ref{fig:R1}, that at increasing of a delay time  two new
lines become visible in the phonon excitation spectra. One of them
($\lambda = 6933,11~\text{\AA}$) is some more to the up of the
$R_{1}$ line, and another, very intensive ($\lambda =
6934,3~\text{\AA}$) is some more to the down. From the different
literary data follows, that the level $4I3$ or practically
coincides with $R_{1}$-line~\cite{Kis69}, or lays above on
0,5~$cm^{-1}$~\cite{kan70,mont87}. At us the difference between
$\lambda = 6933,11~\text{\AA}$ and $R_{1}$ lines is almost
1,0~$cm^{-1}$. Besides, as against $4G3$ and $4H3$ lines, this
line is not observed at 0,5~$\mu$s delay time. Therefore for the
statement, that is the line of pairs $4I3$ there are no major
bases. The not clearest is occurrence of phonon peak with $\lambda
= 6934,3~\text{\AA}$. At delay times less than 1,0~$\mu$s it is
not observed at all, and then very fast grows with increasing of
delay time, i.e., there is an effective phonon generation delayed
in time at this wave length. As  the time dependence of this
phonon peak sharply differs from ones of $4G3$ and $4H3$ (and also
the third and distant neighbors) we count, that this line
originate from strongly perturbed $Cr$ single ions, near to but
not part of  pairs. Indirect confirmation of this conclusion is
that almost on the same distance from a $R_2$ line the phonon peak
(see Fig.~\ref{fig:R2}) also is observed. As you can see in this
Figure the dip in the backward photon echo amplitude is by to the
strong resonant absorption of the  second  laser pulse, twice past
through a sample. It is visible, that the  pairs $4H3$ does not
render practically any influence on a photon echo signal. Against
this some influence of new centers  with $\lambda =
6933,13~\text{\AA}$ and $\lambda = 6934,3~\text{\AA}$ affecting on
the photon echo signal shape of uprise and wane versus wave
length. It, in our opinion, again testifies for the benefit of
that these two lines belong to single chromium ions. Qualitatively
similar to our phonon excitation spectrums the spectra of
fluorescence excitation were observed in~\cite{selz78} at
5,0-$\mu$s delay in ${0,51~at\%~ Cr^{3+}}$~ruby at 6~K  in the
region of the $R_{1}$~-~$R_{2}$ lines. It was assumed in
~\cite{selz78}, that the new lines of fluorescence excitation in
vicinity of the $R$ lines, are originate from $R'$ ions.
In~\cite{selz78} the lines similar to $6920,17~\text{\AA}$ and
$6934,13~\text{\AA}$ were concerned to $N_2$-excitation lines, and
similar to $6933,13~\text{\AA}$ was concerned to $N_1$-excitation
line.

Thus, it is shown, that the time-resolved photo-phonon
spectroscopy appears very sensing method of detection of optical
transitions in those excited states which have the strong
nonradiative relaxation. A time resolution of the method is
defined by a bolometer risetime (it is about $30,0~ns$ in the case
of the $In$-bolometer and $20,0~ns$ in the case of the
$Sn$-bolometer) and allows to observe the spectra evolution on a
nanosecond time scale. A combination of this method and detection
of the time-resolved fluorescence excitation spectra give more the
complete physical pattern of such phenomena, as, for example,
process of energy transfer between single ions and pairs of the
nearest neighbors, so up to the end not solved to this day, by
that there are all new and new models~\cite{jam97} at this
problem.

The authors would like to thank Dr. N.K.Solovarov for the helpful
discussion of results. This work was in parts support of RFBR
grants No. 00-02-16510, 01-02-17730, 02-02-17622 and, ISTC grant
No.~2121.

\bibliography{vlisin}

\end{document}